# Wakefields Generated by Electron Beams Passing Through a Waveguide Loaded With an Active Medium

Andrey Tyukhtin[1], Alexei Kanareykin[2], Paul Schoessow[2]

[1]*Radiophysics Dep. of St.Petersburg Univ., 1 Ul'yanovskaya, St.Petersburg, 198504, Russia*
[2]*Euclid Techlabs LLC, 5900 Harper Rd., Solon OH 44139, USA*

**Abstract.** The wakefields of a relativistic electron beam passing through a waveguide loaded with an active medium with weak resonant dispersion have been considered. For the calculations in this paper the parameters of the medium are those of a solution of fullerene ($C_{60}$) in a nematic liquid crystal that exhibits activity in the X-band [4]. It was shown that several of the TM accelerating modes can be amplified for the geometries under consideration; structures in which higher order modes are amplified exhibit essential advantages as PASERs. In particular, the amplification of the highest mode occurs in a structure loaded with a rather thick active medium layer that maximizes the energy stored by the active medium.

## INTRODUCTION

The possibility of using an active medium to amplify the generated wakefield of a beam and employing the amplified wakefield to accelerate a second beam has been recognized recently. This acceleration scheme is one of several related methods referred to as Particle Acceleration by Stimulated Emission of Radiation (PASER), where an active medium is used to provide the energy for accelerating charged particles. Initial theoretical work in this area focused on acceleration in gaseous $CO_2$ and ammonia laser media [1, 2]. Recently a new active material operating in the X-band has been proposed: a solution of fullerene ($C_{60}$) in a nematic liquid crystal has been found to exhibit a maser transition in the X-band frequency range [3, 4]. The ability to employ a microwave frequency material simplifies the construction of test structures and facilitates beam experiments using Dielectric Loaded Accelerator (DLA) concepts [4].

We present here the results of analytical and numerical studies of the Cherenkov radiation from a Gaussian electron bunch moving through an accelerating structure loaded with an active medium. The analysis of the problem is based on the waveguide mode formalism applied to a resonant dispersive medium excited by external optical pumping. The theory of Cherenkov radiation in waveguides loaded with a dielectric medium was outlined previously in variety of publications (see, for example, the review article [5]). However, Cherenkov radiation in waveguides with dispersive media (passive or active) has not been extensively discussed in the literature. For example, in Ref. [6] the frequency dependence of the conductivity is taken into account. Some papers are devoted to the investigation of radiation in a waveguide completely or partly filled with a passive dielectric exhibiting resonant dispersion [7,8]. The case of an active medium with strong resonant dispersion is considered in the paper [1]. Further expansion of the theory can be found in Ref. [2].

## SOME COMMON RESULTS

The bunch is assumed to move along the axis of a vacuum channel in a cylindrical waveguide with an active media layer. The charge density is determined by the expression $\rho(x,y,\zeta)=\frac{q}{\sqrt{2\pi}\sigma}\exp\left(-\frac{\zeta^2}{2\sigma^2}\right)\delta(x)\delta(y)$, where $\zeta = z - Vt$ is the distance to the bunch, $q$ is the charge of the bunch, and $\sigma$ is a parameter characterizing the bunch length. The radius of the bunch is assumed to be negligible. The geometry of the problem is shown in Fig. 1, where $a$ is a waveguide radius, $b$ is the radius of the vacuum channel, and $d = a - b$ is the thickness of the active medium layer.

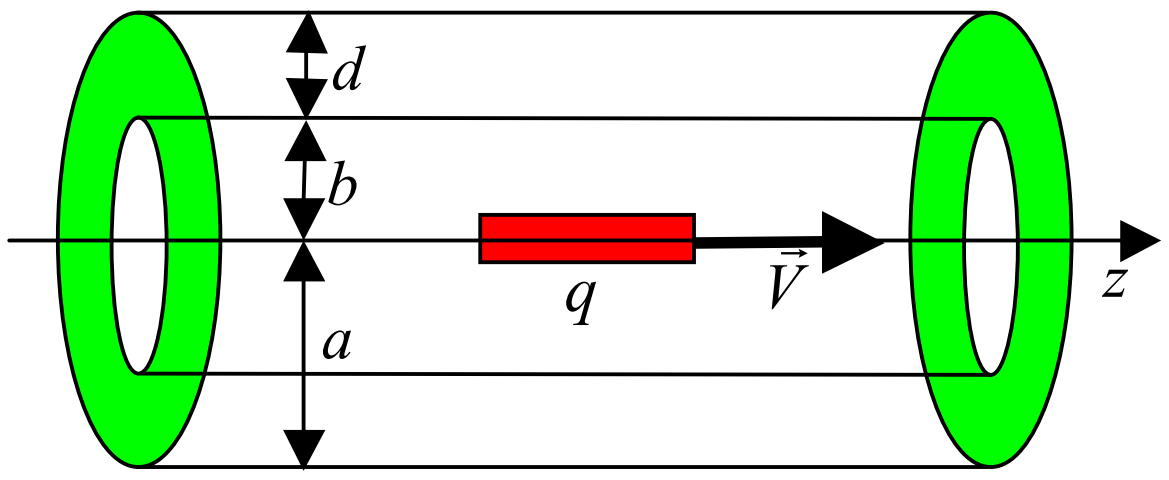

**FIGURE 1.** Geometry of the problem.

The common results for the analogous problem with a passive dispersive medium were included, for example, in Ref. [5]. It can be shown that these results may be generalized for the case of an active medium. We obtain the following expression for the longitudinal component of the electric field at a distance greater than the bunch length:

$$E_z = -\frac{4q\left(1-\beta^2\right)}{c^2\beta^2}\sum_{m=1}^{\infty}\sum_{j=1}^{2}\mathrm{Re}\left\{\frac{\omega[s\,K_1(kb)\psi_1(s)+\varepsilon\,k\,K_0(kb)\psi_0(s)]}{\frac{d}{d\omega}[s\,I_1(kb)\psi_1(s)-\varepsilon\,k\,I_0(kb)\psi_0(s)]}\times\right. \tag{1}$$

$$\left.\times I_0(k\,r)\exp\left(-\frac{\omega^2\sigma^2}{2V^2}\right)\exp\left[i\omega\left(\frac{z}{c}-t\right)\right]\Bigg|_{\substack{\omega=\omega_{jm},\,k=k_{jm}\\ s=s_m}}\right\},$$

where $r=\sqrt{x^2+y^2}$ is the distance from the waveguide axis, $\beta = V/c$, $\psi_0(s) = J_1(sb)N_0(sa) - J_0(sa)N_1(sb)$, $\psi_1(s) = J_0(sb)N_0(sa) - J_0(sa)N_0(sb)$, $k(\omega)=\omega V^{-1}\sqrt{1-\beta^2}$, $k_{jm}=k(\omega_{jm})$. Here $J_n(\xi)$ are the Bessel functions, $N_n(\xi)$ are the Neumann functions, $I_n(\xi)$ are the modified Bessel functions, and $K_n(\xi)$ are the modified Hankel functions. The introduction of the summation over $j = 1,2$ is convenient because in the case of the medium considered below each value of $s_m$ corresponds to two frequencies $\omega_{1,2m}$. The frequencies of the harmonics are determined by the following system of equations:

$$s(\omega) = \omega (c\beta)^{-1} \sqrt{\varepsilon(\omega)\mu(\omega)\beta^2 - 1}\,, \qquad (2)$$

$$s(\omega)\, I_1(k(\omega)b)\psi_1(s(\omega)) - \varepsilon(\omega)\, k(\omega)\, I_0(k(\omega)b)\psi_0(s(\omega)) = 0\,. \qquad (3)$$

## THE CASE OF WEAK RESONANT DISPERSION

The results (1) – (3) are true both for passive and for active medium with arbitrary dispersion characteristics. We consider here a typical case when the dispersion has a resonant character and there is only one resonant frequency in the range of interest for us. In this case the refractive index has the form $n^2 = n_{c0}^2 + \dfrac{\omega_p^2}{\omega_r^2 - \omega^2 - 2i\,\omega_d\omega}$, where $\omega_r = 2\pi\nu_r$ is the resonant frequency, and $\omega_p = 2\pi\nu_p$ is a plasma frequency. For a passive medium $\omega_p^2 > 0$, but for an active medium $\omega_p^2 < 0$, i.e. $\omega_p$ is imaginary. In the first (passive medium) case the dispersion is normal almost everywhere except in the vicinity of the resonance but is anomalous in the neighborhood of the resonance. In the second (active) case the dispersion is anomalous almost everywhere except in the resonance vicinity and is normal in the resonance vicinity where the imaginary part is negative.

The basic thrust of this paper is the analysis of the active medium case, where the dispersion is relatively weak and energy losses are relatively small (the opposite case of strong dispersion was analyzed in [1, 2]). These conditions are fulfilled if the following inequalities are satisfied: $\left|\nu_p^2\right| << \nu_r\nu_d$, $\operatorname{Im} n_{c0}^2 << \operatorname{Re} n_{c0}^2$. Analytical and numerical calculations showed that in the given situation complex frequencies of harmonics of the first series ($j = 1$ in (1)) are located in the vicinity of the resonant frequency and possess negative imaginary parts. The amplitudes of these harmonics decrease exponentially ($\exp(-\omega_d t)$). As a rule these harmonics excited by the beam exhibit relatively weak magnitude, and therefore their impact is negligible. Usually the frequencies and amplitudes arising from second series of harmonics ($j = 2$ in (1)) are close to one for the dispersion-free case, i.e. $\omega_{2m} \approx c\, s_m / n_{c0}$. However this condition occurs if all frequencies of the second series are not close to the resonant frequency. If some mode frequency $\omega_{2M}$ is in the vicinity of the resonant frequency, its complex frequency is

$$\omega_{2M} \approx \omega_r + 0.5 \cdot i \cdot \left[ \sqrt{(\omega_d - \chi\omega_r)^2 + \left|\omega_p^2\right| / \operatorname{Re} n_c^2} - (\omega_d + \chi\omega_r) \right], \qquad (4)$$

where $n_c = \sqrt{n_{c0}^2 - \beta^{-2}}$ , $\chi = \operatorname{Im} n_c^2 / \left(2 \operatorname{Re} n_c^2\right)$. The amplitude of this mode can be decreased or increased depending on the parameters. The amplification condition is $\operatorname{Im} \omega_{M2} > 0$ which may be also written as $\left|\omega_p^2\right| > 2\omega_d \omega_r \operatorname{Im} n_c^2$.

In the low GHz-frequency range, an example of an active medium with weak dispersion is a solution of fullerene ($C_{60}$) in a liquid crystal under optical pumping and an external dc magnetic field. This medium has been discussed as a possible "amplifier" for electromagnetic fields in Refs. [3,4]. In this medium, the amplification process is determined by the magnetic permeability. The dielectric permittivity is nearly constant and its imaginary part determines the dielectric losses of the structure. A model of medium is given by the formulas

$$\mu = \mu_{c0} - \frac{\left|\tilde{\nu}_p^2\right|}{\nu_r^2 - \nu^2 - 2i\nu_d \nu}; \qquad \varepsilon = const; \qquad n^2 = n_{c0}^2 - \frac{\left|\nu_p^2\right|}{\nu_r^2 - \nu^2 - 2i\nu_d \nu}, \tag{5}$$

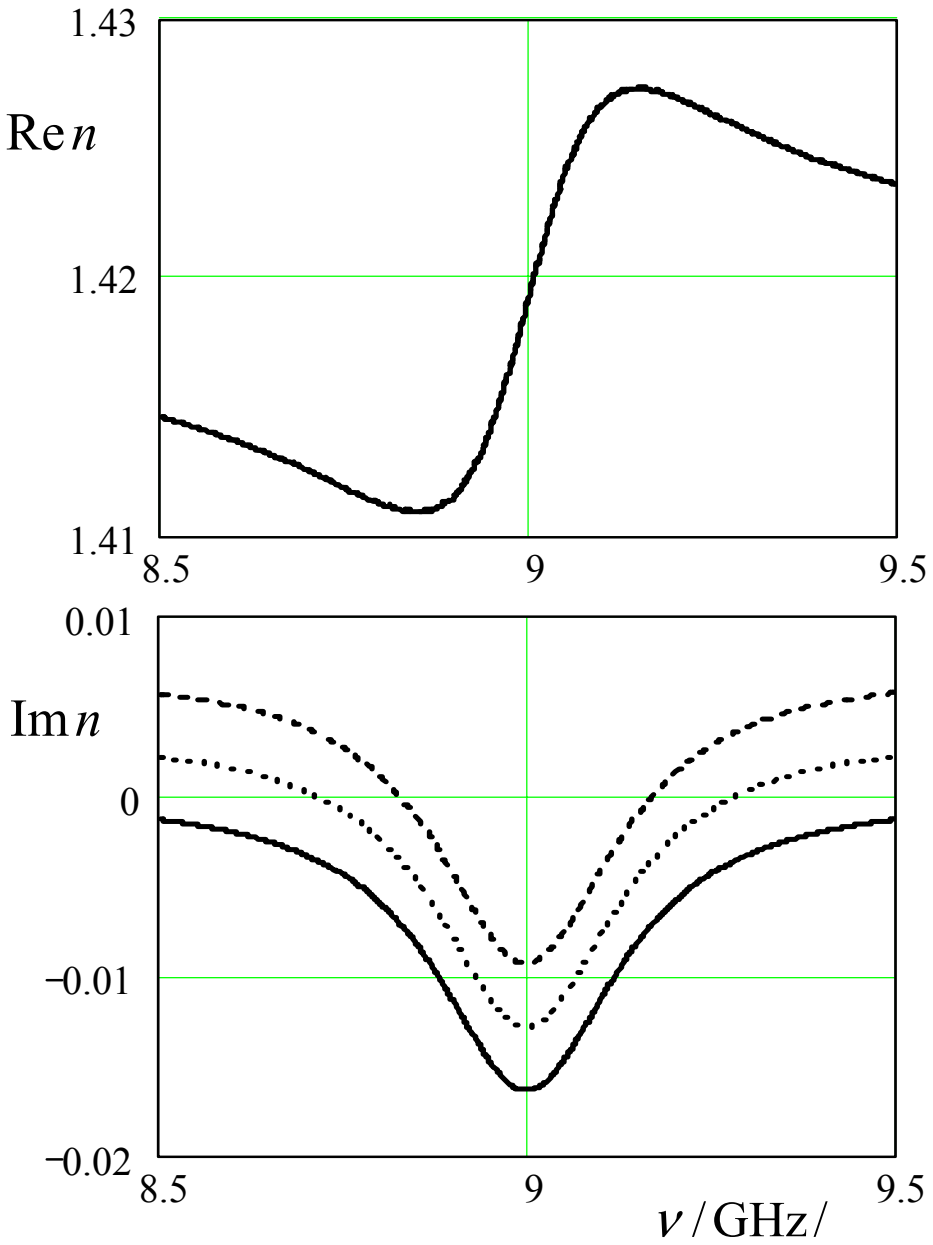

**FIGURE 2.** Real and imaginary parts of the refractive index. Solid, dotted and dashed curves correspond to $\operatorname{Im}\varepsilon = 0, 0.01$, and $0.02$ respectively.

where $n_{c0}^2 = \varepsilon \mu_{c0}$, $\left|\nu_p^2\right| = \varepsilon \left|\tilde{\nu}_p^2\right|$. On the data obtained from electron paramagnetic resonance measurements, the parameters of this material are roughly estimated to be [4]: $\mu_{c0} = 1$, $\nu_r = 9 \times 10^9$, $\left|\tilde{\nu}_p\right| = 250 \cdot 10^6$, $\nu_d = 150 \cdot 10^6$, $\operatorname{Re}\varepsilon = 2$. The imaginary part of permittivity can be determined within the range from 0.005 to 0.02. The real and imaginary parts of the refractive index in the resonance vicinity are shown in Fig. 2.

We next describe some numerical results for the case of an ultra-relativistic beam ($\gamma = 22.37, \beta = 0.999$). First we consider the case of a relatively thin waveguide with a radius of 2 cm to allow the demonstration of the single mode amplification regime for the $TM_{01}$ mode. Fig. 3 shows the dependence of the imaginary and real parts of the frequency on the channel radius for different values of the imaginary part of the permittivity. Note that for $\operatorname{Im} \nu_{21}$, only the region of positive magnitudes is shown. We observe that the maximum amplification and the range of channel radii where the amplification takes place decrease as the imaginary part of the permittivity increases. However this decrease is not very significant when $\operatorname{Im}\varepsilon < 0.02$.

The wakefield for the case of maximum amplification is shown in Fig. 4. From now on we assume that the beam is relatively short ($\sigma = 1\,\text{mm}$) and the total charge in the bunch is $q = -1\,\text{nC}$. Note that the e-folding amplification distance for the first mode is $l_1 = 63.2\,\text{cm}$ for the case of $\operatorname{Im}\varepsilon = 0.02$, and $l_1 = 40.7\,\text{cm}$ for $\operatorname{Im}\varepsilon = 0$.

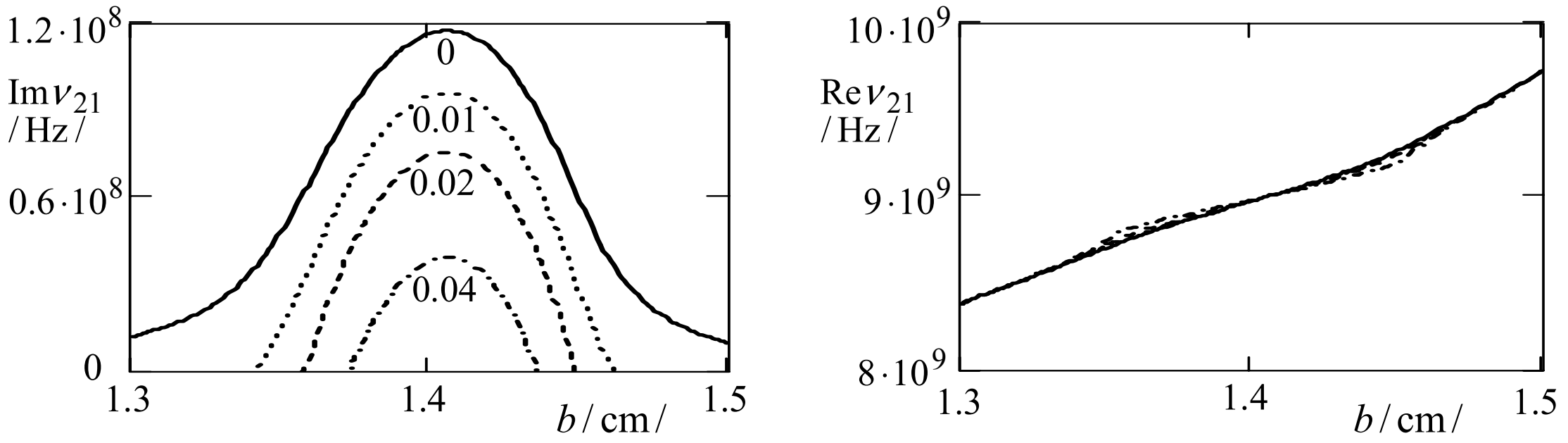

**FIGURE 3.** Dependence of the imaginary and real parts of the mode frequency on the channel radius for $a = 2\,\text{cm}$. The corresponding values of $\text{Im}\,\varepsilon$ are indicated.

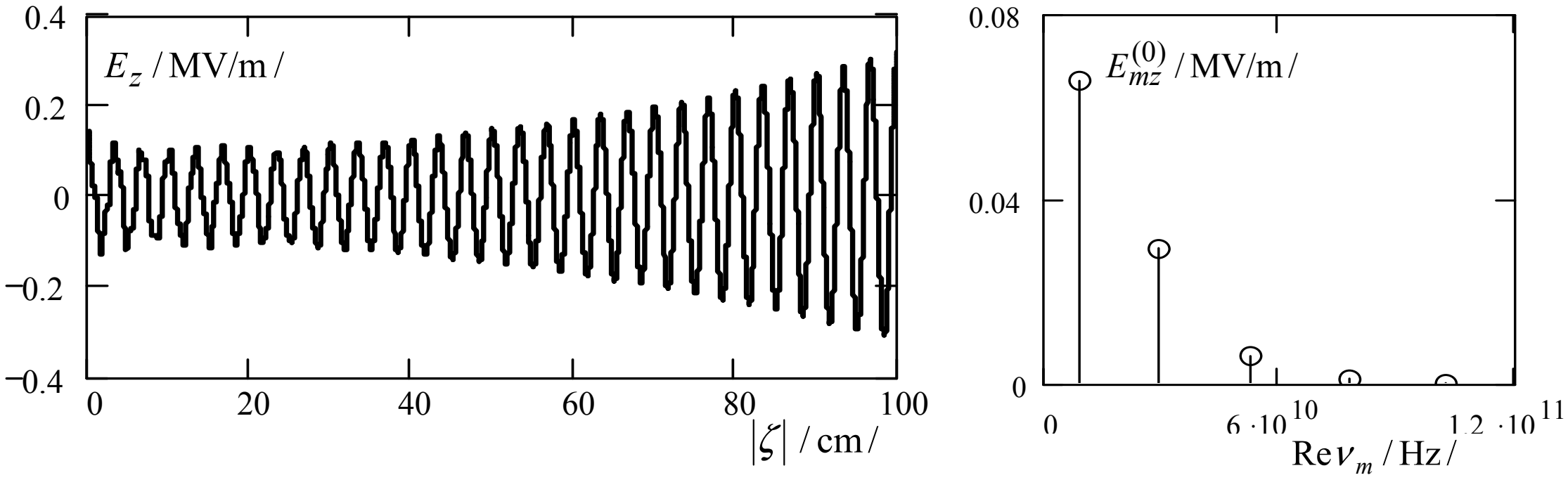

**FIGURE 4.** Longitudinal component of the electric field $E_z$ as a function of the distance $\zeta = z - Vt$ behind the bunch. (left) The amplitudes of the harmonics $E_{mz}^{(0)}$ at $\zeta = 0$. (right) ($\sigma = 1\,\text{mm}$, $q = -1\,\text{nC}$, γ=22.37, $\beta = 0.999$, $\text{Im}\,\varepsilon = 0.02$, $a = 2cm$, $b = 1.405cm$)

Next we consider the case of relatively thick waveguide with $a$=10 cm, in which the amplification regime for several modes can be found. As we see from Fig. 5, one can obtain the amplification of the modes with numbers from 1 to 6 depending on the channel radius. The maximum values of $\text{Im}\,\omega_{2m}$ depend on $m$ rather weakly, and one can use a set of modes for wakefield amplification. Meanwhile the amplification regime for a single mode occurs only for the case of a very thin layer of active medium. The amplification is very sensitive to the geometric parameters of the structure (see curve 1 in Fig. 5 which is rather narrow). The amplification regime for the higher modes occurs in structures with thicker active medium layer. The thickest active medium layer corresponds to amplification of the 6th mode. This has the important advantage that the increased volume of active medium stores more energy inside the structure. The amplification of the higher modes is also not very sensitive to the exact choice of channel radius (the curve for the 6th mode is rather wide).

Figure 6 shows the wakefields in the case of maximum amplification of the 1, 4, and 6 modes. The channel radii and e-folding amplification distances of the corresponding modes are indicated. The right hand plot in Fig. 6 gives the corresponding spectra at the distance $\zeta = 0$. The single-mode wakefield is excited in the case of a very thin medium layer. In the other cases the wakefields are multimode.

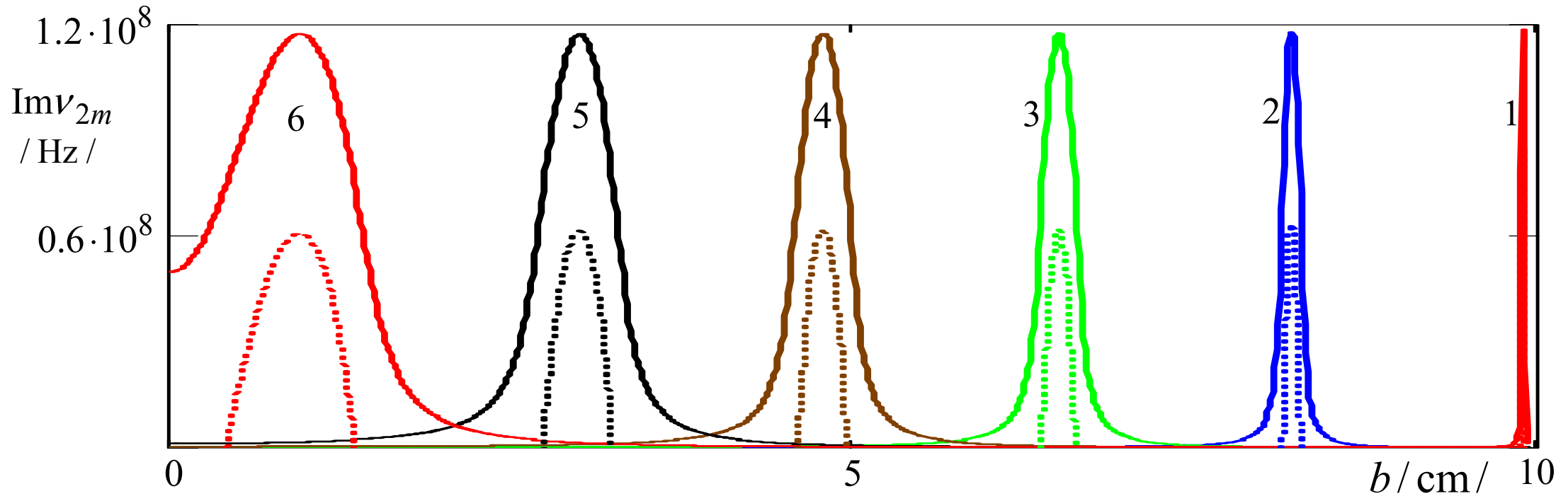

**FIGURE 5.** Dependence of the imaginary parts of the modes frequencies on the channel radius for $a = 10\,cm$. Solid and dotted curves correspond to $\mathrm{Im}\,\varepsilon = 0$ and $\mathrm{Im}\,\varepsilon = 0.02$ respectively. The mode number corresponding to each curve is indicated.

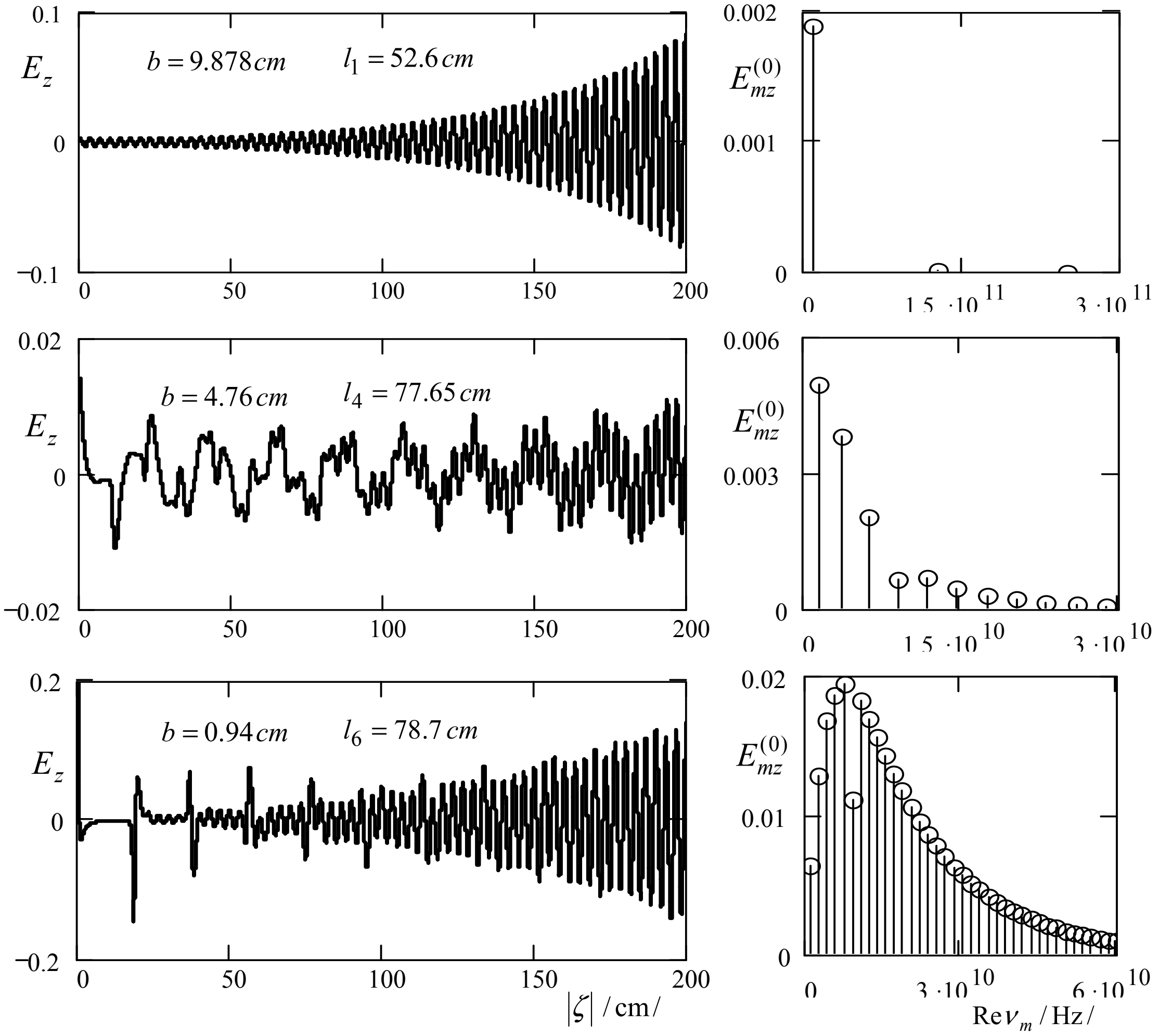

**FIGURE 6.** Longitudinal component of electric field $E_z$ as a function of the distance from the bunch $\zeta = z - Vt$ and the amplitudes of the harmonics $E_{mz}^{(0)}$ at $\zeta = 0$ ($\mathrm{Im}\,\varepsilon = 0.02$, $a = 10\,cm$).

It is interesting that for a distance of $|\zeta| = 200\,\text{cm}$ the amplification of the 6th mode is the most effective provided this mode can also be excited effectively.

## CONCLUSION

In this paper a general analysis of the amplification of wakefields in cylindrical structures by weakly dispersive active media is presented. The wakefields are generated by a relativistic electron beam passing through an accelerating structure loaded with an active dielectric medium [1,2,4]. The parameters of the recently studied active solution of fullerene ($C_{60}$) in a nematic liquid crystal operating in the X-band [4] have been assumed for these calculations. It was shown that in the case of a medium with weak dispersion the amplification effect is determined by the imaginary part of the refractive index. In contrast to this, for the case of medium with strong dispersion the amplification effect is determined by the anomalous character of the dispersion and the contribution of the imaginary part of refractive index was less significant [1,2]. The amplification of several modes has been demonstrated, and the dependence of the maximum increment on the mode number was found to be relatively small. The waveguide geometry determines the amplification conditions for a given mode. The amplification of higher order modes exhibits noticeable advantages: (1) amplification of the higher modes occurs when an electron beam traverses a structure loaded with a thick active medium layer and hence more energy is available from the medium; (2) the dependence of the amplification magnitude on the geometrical parameters of the structure for the higher modes is not very strong; (3) as a rule, the excitation of the highest amplified mode is rather effective.

## ACKNOWLEDGMENTS

We would like to thank L. Schächter for his help in our studies, and W. Gai and J.G. Power for useful discussions. This work was supported by the US Department of Energy, Division of High Energy Physics Grant # DE-FG02-05ER84355 and by the Russian Foundation for Basic Research, RFBR Grant # 06-02-16442-a.